\magnification=1200\hfill{SWAT-TH/03-373}
\bigskip
\noindent  {\bf Paul Dirac and the
 pervasiveness of his thinking}
\vskip 30pt
\noindent {\it Invited Lecture presented at the Dirac Centenary Conference,
hosted by
Baylor University, Texas;  30th September-2nd October 2002}
\vskip 40pt
\medskip{\bf David I Olive}
\medskip
\noindent{\it Physics Department, University of Wales Swansea, Swansea SA2 8PP, UK}
\vskip40pt
 I shall use a few personal reminiscences of my time 
as a student and
 colleague of Dirac in Cambridge to introduce 
 some reflections 
on the nature of research in theoretical physics. 
I shall discuss
 and illustrate the
approach of Dirac to his own research
 and the pervasiveness of the  influence 
his example has provided.

I shall discuss how  ideas produced at all stages of his career
 have proved
 to be extraordinarily visionary, still motivating
 and exerting an influence
on research many years after his death. 
Examples include his celebrated equation
and prediction of 
the existence of antiparticles, the magnetic monopole and 
its quantisation condition, the quantum theory of dynamical systems
with constraints and the membrane theory of the electron.

\medskip
\vskip250pt
\vfill

I would like to start by expressing my gratitude 
to the organisers for inviting
 me to this fine conference
 commemorating one of the greatest figures in
 twentieth 
century physics. I have to admit that the first time I came 
across
 the name of Dirac was when, as a boy, I read a science 
fiction story about
 a spaceship powered by a \lq\lq Dirac drive''.
 I was given the impression
 that the \lq\lq inventor'' of this
 was someone extraordinary but I did not then realise 
in what way.
 I have never been able to identify the author or title of
 this story subsequently.

Many years later, when I came
 to the University of Cambridge as an affiliated student
 to study for the
 mathematics tripos, chance would enroll me at 
St John's College,
 where Dirac was a professorial fellow.
 By design, I attended 
the two lecture courses he gave in
 the academic year 1959-1960, 
quantum mechanics and advanced quantum mechanics.
The next year
 I became a research student and yet later
 a research fellow 
and faculty member in DAMTP. 
So latterly, and for the four years up until
 his retirement in 1969,
I became a colleague of Dirac, albeit 
a very junior one,
 and with little personal contact
 but enough to realise 
that the commonly accepted picture of
 him as pathologically 
shy and monosyllabic was untrue and unfair.

Like many others, 
I found his lectures to be a revelation, at last laying out
the previously mysterious subject of quantum theory in
 a comprehensible way, with honesty, beauty and clarity.
 That year I started attending the weekly seminars
 in theoretical physics
and one of the privileges
 was to see Dirac take his seat in the front row of the audience.
 Later, as the seminar progressed,
 he could often be seen gently
 lowering his forehead onto
 the desk in front of him. It was even
 whispered that he was
 out of touch with modern developments.
To some extent this was true and I think that he was not interested
 in
 the analytic properties of the scattering matrix nor
the group theoretic classification of particle states,
topics which were then so much the centre of attention.
 Of course what he was doing
 was simply to follow his own 
ideas and his own sense of priorities which 
were illustrated 
by the subject matter of his advanced lecture course 
and by
 seminars he  himself gave from time to time.
 Examples of 
the latter concerned \lq\lq The bubble theory of the electron'',
 \lq\lq A
remarkable representation of the 3+2 de Sitter group'' and
\lq\lq Quantum electrodynamics without dead wood''.
 The first two of these were not mainstream topics at the time but now 
we are able
 to see how prescient these interests
 were and how much he 
anticipated lines of research
which were taken up
thirty-five years later in the theory of superstrings and branes, 
 a subject  now very fashionable and acknowledged to be the most promising 
candidate for a truly unified theory. The third topic seemed
 reactionary 
 but again has, to a large extent, been vindicated by later developments.

The research groups that made up DAMTP were housed together
for the first time
in 1967 or 1968
when premises between Silver Street
 and Mill Lane were
 vacated by the University Press,
 but Dirac was not assigned an office there as he 
apparently preferred 
to work at home. Nevertheless
 he made a point of attending
 the morning
 coffee break 
once a week in order to distribute preprints he had received 
and to update us on any relevant gossip. For example, I remember
him breaking the tragic news of the death of the Swedish theoretical
physicist Gunnar K\"all\'en in the crash of his own self-piloted plane.

The only time that I shared a meal with Dirac was at the Director's 
table
in the restaurant in ICTP in Trieste during a conference
 on \lq\lq 
Renormalisation Theory'' in August 1969. 
On that occasion he was rather communicative 
and keen 
to describe an interesting book that he was reading 
concerning
 the life and death of Napoleon 
(It may be relevant  that Dirac's family
 originated 
in the Charentes region of eastern France with only a
 few 
generations living in Switzerland before settling in England).
 He was particularly
 intrigued by 
the arsenic poisoning that led to Napoleon's
 death in exile on the
 island of St Helena and the theory
 propounded by the book that
 this was part of a systematic 
plot on the part of someone.
 I have since heard of an 
alternative to this conspiracy theory
 whereby the arsenic
 can be traced to the Paris green dye in the fashionable
 wallpaper that still hangs in his bedroom there.
This dye
 could not cope with the tropical climate and arsenic
leached into the atmosphere.

These minor personal contacts that I had with
Dirac indicate to me that he was willing to 
reveal a very
 human side to his character. In my
 experience as a 
young man at that time this
 was not so common amongst
 senior British academics.

I saw Dirac only once after the lunch in Trieste and I cannot resist
mentioning this as the occasion was so strange. About ten years after the 
Trieste encounter
I was eating in the restaurant of CERN, which, as you can imagine, was large and crowded, when suddenly at a distance Dirac appeared,
 unheralded and unannounced, (escorted by J Mehra), 
and just as suddenly disappeared.

Each year that he was in Cambridge,
 Dirac gave his
 lecture course on quantum mechanics based on topics 
selected from his book, occasionally supplemented by
 a more advanced
 course including the remaining topics
 and also his theory of 
constrained dynamical systems and their quantisation.

 Since Dirac seemed
 to play little role in departmental
 affairs he could 
only exert 
an influence by  example,
the example of his exposition of quantum mechanics  and that
of his many research papers.
In all this he demonstrated how 
it is possible to make every detail 
of an argument visible
 and clear in a well-balanced way.
 Pursuit of logic was a key factor.
 Followed ruthlessly 
it led Dirac to the most amazing conclusions
 and insights.

 We all know that he put the finishing touches on 
the theory 
and formulation of quantum mechanics adding to its 
coherence. 
But more importantly this work provided the foundations for
his pioneering work in taking
 the next steps beyond quantum
mechanics to take into account the structure of space-time.
He considered the effects
  of special relativity, what is vaguely called locality,
 general covariance and even some global topology.
 I shall not be able to describe his most celebrated achievement, the
Dirac equation,
 what he liked to call \lq\lq the 
relativistic wave equation
 for the electron", [Dirac 1928] nor the subsequent
 prediction of antiparticles [Dirac 1931]
(I have reviewed this aspect of Dirac's work  elsewhere [Olive 1997]). 
Nor shall I be able to say
much about his procedure of \lq\lq second quantisation" which
relates the principles of quantum mechanics to the structure of space-time
and establishes a correspondence between fields and particles [Dirac 1927].
 But
 I do want to to say more about another way in which
  the notion
of locality was developed and I shall do this later on.

There are many ideas about scientific method and the
 philosophy behind different approaches. 
The truth 
is that the success of a method 
will depend both on the
subject matter and 
the particular scientific worker. 
It is also
 true that different methods work better
at different stages in the  development 
of the same  subject.

 Dirac pursued what might be called the aesthetic approach 
and
in his later writing he  emphasised the courage required
 for this
 sort of work, the courage to pursue an argument to
 its remorseless
 conclusion despite the fear that a sudden 
quirk
 could well spell disaster. This sort of
 work requires
 other qualities that he did not emphasise, for example, a
searing honesty, that he so transparently did possess.
 Besides this,
 success also requires hard work and dedication,
 not to mention the
 sheer genius that no
 one can explain.

It has to be admitted that the idea of beauty as a criterion
 for scientific 
truth is fraught with danger as it can be too
 subjective and can easily
 fail in the wrong hands. But it
 is important to realise that the beauty
 concerned is not just
 a snapshot, it also has a quality of overall
 coherence which
 I think is important. In the hands of Einstein and
 Dirac and 
others the aesthetic approach led to the founding of the cornerstones
of physics as known in the first half of the last century.

The initiation of a new era in physics  in the middle of the
 last century
was  one of the side effects of World War II.
Perhaps the main effect was in applied physics
 but fundamental
 physics was affected too. Previously,
Einstein and Dirac 
 had concentrated their
 thoughts on the electromagnetic 
and gravitational
 fields so important on the macroscopic level.
But the frightful effectiveness of the atomic bomb made it clear,
even to the general public, that there were other important fields
in nature, namely those associated with the structure of the nuclei
of atoms. Postwar, new particle accelerators were built to
 investigate 
this,  and they provided a cornucopia of many,
 unexpected new elementary
 particles together with information 
concerning their properties.
 Now the aesthetic approach waned 
in importance as a more pragmatic
 approach was required to succeed in
 unravelling the patterns hidden in this data.

This process took about thirty-five years 
and in the early 1980's culminated in
what is
 called the \lq\lq standard model''. 
This mundane title disguises
the fact that this model
 synthesises a number of extraordinary features
that 
I shall now briefly describe.
The model is formulated
 in terms of a choice of a number of particles
 (or equivalently their corresponding fields) together with 
the specification of a system of 
the non-linear coupled 
partial differential equations of motion
that these fields
 satisfy
(or, more succinctly, by a Lagrangian).

The first surprise is the indirect nature
of the good choice 
of ingredient particles
and it is this that partly explains the 
length of 
the period
 of gestation.
 Neither the proton nor neutron nor any of
 the newly discovered
 post-war particles participate. 
Instead there are 
two classes in the choice,
one augmenting
 the electron and the other the photon 
(the quantum of the electromagnetic field). 
These are the two particles most familiar to prewar physics.
The electron class contains the neutrino and 
strongly interacting counterparts of the 
electrons called \lq\lq quarks'' out of which
the proton and neutron
 can, in a sense, 
be constructed. The photon 
is augmented by
 other similar particles of unit 
spin, namely the three
 weak gauge particles $W^{\pm}$ and $Z^0$
 and 
the eight colour gluons
which couple to themselves 
and the quarks forming 
a subtheory called QCD
 (responsible for strong forces).
Out of these extras only the electroweak gauge particles
are observed directly
(originally at CERN). The quarks are 
observed only indirectly 
(originally at SLAC) and the colour 
gluons even less directly.
Actually there are two extra copies
of the electron class 
(called generations)
 involving the muon and the tau-lepton respectively, 
together 
with extra neutrinos and quarks that occur as constituents
of
 the  particles that were discovered postwar.
Another
ingredient is a set of scalar particles, still being sought,
whose fields (the Higgs'), unlike all the others, fail to vanish in the vacuum.
The effect of this is what is called a \lq\lq spontaneous
breaking" of the gauge symmetry and a mechanism for providing
mass in a natural way.

The second surprise was that, given the good choice 
of fundamental fields
just described, the basic equations
 that they satisfy are found simply 
to be elaborations of
 the familiar equations of Maxwell and Dirac 
that were already known to describe electrodynamics. 
Indeed a treatment of
this system 
of equations 
 appears in later editions of Dirac's text book
 on quantum mechanics
and was included in the
 subject matter of his advanced lecture course.

Maybe, after all, it is possible to understand why Dirac used to
  lower
his forehead ever so gently onto his desk. 
Unfortunately the death of Dirac
roughly coincided
with the acceptance of the standard model within the particle physics
community. Judging by his later writings he remained
unhappy with postwar developments in quantum electrodynamics
and apparently unaware of the standard model and the extent to which
it embodied ideas more familiar to him.
 But we do know
that he was pleased with the idea 
of one of the important ingredients,
the quark model of hadrons. 
In 1966 Murray Gell-Mann, its proposer,
visited Cambridge in order to spend a few months as an Overseas Fellow
at Churchill College.
 On his arrival
 he went to pay his respects to Dirac 
and returned reporting that Dirac was 
delighted to hear of
 the role his relativistic wave equation was
 likely to play
in view of the fact that electrons 
and quarks have the same spin. In retrospect it is clear
that the introduction of the quark model marked the turning point
back towards the Maxwell-Dirac theory in the history of the 
development of the standard model. Yet neither it, nor the
unified electro-weak gauge theory of a few years later, were
met with automatic acceptance by the sceptical particle physics community.
Indeed Werner Heisenberg, for example, was quite
antagonistic to
 the quark model, presumably because of his own
 rival theory,
the non-linear spinor theory.
 I remember being taken aback
when he publically
 reprimanded me in June 1971 for trying to justify 
string theory
as a relativistic version of the 
quark model. \lq\lq The quark
 model is not physics" 
were his words.

Of course there is a lot of devil in the detail covered
 by the word
\lq\lq elaboration" used above with
 reference to the Maxwell-Dirac
system of equations
 and I shall say something about it.
 But the fact 
remains that
physics had turned full circle. 
The reason was that, underlying
all the complexity
 of the particle accelerator data, is a remarkably
simple underlying principle that had been known in
 its embryonic form
since its enunciation in 1929.
 Despite this, it  had initially met
 with enormous
 scepticism
when proposed in the wider context of the
 postwar data. Of course,
I am talking
 about \lq\lq the gauge principle"
  a notion that exploits
the idea of
 locality that I have already promised to explain.

Before doing this I want to say how the gauge principle 
applied
 to the standard model brings into play something
new compared to the earlier version. This is a branch
 of pure mathematics
that is tailor-made to describe
 all the details
 neatly and succinctly and is
known
 as the theory of compact Lie groups and their representations.
The particular Lie groups occurring in the standard model are
 straightforward
matrix groups called $U(2)$ and $SU(3)$ referring
 to the electroweak and strong forces respectively. 
These groups are formed out of unitary matrices
with two 
and three rows and columns and hence are continuous
 and 
have dimensions 
$2\times2=3+1$ and $3\times3-1=8$ respectively,
 equalling the number 
of gauge particles already mentioned when
 the photon is included.
 The quarks and leptons
making up what
 was called the electron class in the explanation
 above likewise
transform according to some natural representation of low dimension. 

Thus all the data gathered at such expense and effort 
at particle
accelerators around the world can be
 encapsulated by the gauge principle augmented
by 
a few group theoretical rules. That is why the standard
 model
is such an achievement. 
A plethora of {\it ad hoc} theories,
 four-fermi theory, Yukawa theory, isovector pion theory,
 flavour symmetry, V-A theory and so on, are all subsumed in,
 or superseded by one all encompassing model based on an 
aesthetic principle. The power
 of this revolution is that the evidence 
is provided by data produced by sceptics, 
namely the \lq\lq hard-nosed'' experimental physicists
who
 have no truck with \lq\lq airy-fairy" ideas such
 as the gauge principle.

It is important to realise that
  the principles of quantum mechanics
and relativity remain intact
 and indeed
 are elevated to a larger range of validity whilst 
a new principle is included. The challenge emerges of
 finding an underlying explanation 
of
 the new picture with its rules
and in developing 
the underlying principles further.
 Maybe even gravity
can be incorporated.
This pursuit is roughly known as \lq\lq beyond the standard model'' 
and it sets the stage
for more aesthetic considerations and
 a renaissance 
of so many of Dirac's pioneering ideas.

In order to introduce the work in which Dirac, 
with possibly
his greatest stroke of genius, 
saw how to exploit the gauge principle
to disclose aspects of quantum electrodynamics 
that still tantalise
seventy years later,
 I shall recall the beginning of the story 
of quantum
mechanics.
 It started with the 
radical proposal of Planck, clarified by Einstein, that the energy, $E$,
of electromagnetic radiation of a
 given discrete angular frequency, $\omega$,
occurred in quanta:
$$E=n\hbar\omega,\qquad n=0,1,2,3,4\dots$$
where $\hbar$ is Planck's constant. 
It took almost thirty years of research to formulate principles
 of quantum mechanics sufficently well that this result could 
be derived by applying these principles to Maxwell's equations
which were known to describe the radiation. Dirac was one of those
who helped achieve this with his theory of second quantisation [Dirac 1927].

 An obvious follow-up strategy was to survey natural phenomena
and try to identify similar patterns of integer quantisation.
One striking example was provided by the values of
 the electric charges, $q$, of the
particles known to make up atoms. They were
observed to follow an analogous rule:
$$q=nq_0,\qquad n=0,\pm1,\pm2,\pm3,\dots\eqno(1)$$
This rule  explains why an atom in its natural
un-ionised state is electrically
neutral, a fact that is important for the stability of massive objects such as planets. But it is difficult to explain this result
in terms of the principles of quantum mechanics alone
since it is difficult to imagine a classical system in which 
electric charge is a dynamical variable in the way that the energy
of radiation,  in Planck's relation above,  is. So what could
 explain this pattern, whose
validity survives the discovery of all the new post World War II particles,
and is, in fact, still  the most striking feature of particle physics?
Dirac found an answer in 1931
and  it brought into play not one but
two new ideas outside the framework of
 quantum mechanics, one of which was the gauge principle [Dirac 1931].
But it also brought a disappointment that continues to haunt
theoretical particle physics today.

Now I come to explain what the gauge principle is, 
and something of the idea of locality. 
I have deferred
this until I can no longer avoid 
introducing equations and formulae
and I apologise
 in advance for this. 

 According to Schr\"odinger's
wave mechanical formulation of quantum theory, 
any electrically charged particle, such as an electron,
possesses a wave function $\psi(x)$ that takes complex values.
The argument  $x$ denotes the space-time point 
at which the wave function is evaluated.

The wave function determines what is called the state of the electron,
namely the totality of physical attributes that can be measured,
but an awkward feature is that different 
(normalised) wave functions correspond
to the same state if they differ by an overall phase factor, $e^{i\phi}$,
that is independent of the space-time point, $x$. Thus, briefly,
nothing physical is changed by the substitution:
$$\psi(x)\rightarrow e^{i\phi}\,\psi(x),\qquad \phi\hbox{ constant.}$$

This feature is more disquieting when one remembers that,
 according to special relativity, no information signal can travel faster
than the speed of light which is finite. So, if $x$ and $y$ are two
distinct space-time points at the same time, there is no way that an observer at
$x$ could know that the phase of the wave function had been altered at $y$.
This consideration suggests that it would be more reasonable
to require that the phase change in the substitution above could be allowed 
to vary over space-time:
$$\psi(x)\rightarrow e^{{iq\over\hbar}\chi(x)}\,\psi(x).\eqno(2)$$
So the phase change $q\chi(x)/\hbar$ now varies over space-time and so is said to
be local. $q$ is the electric charge carried by the particle
 whose wave function is $\psi(x)$ and $\hbar$ is Planck's constant, 
both inserted for convenience later on. The gauge principle states that all physical quantities should be unaltered by the substitution (2), which is the gauge transformation of the wave function. As we shall see, an awkward
feature of wave mechanics has been converted into a new principle, 
in fact the most powerful principle of the latter half of the 20th century
as far as particle theory is concerned.

Now the wave function has to satisfy some sort of evolution equation
(such as the Schr\"odinger equation) involving its derivatives
 $\partial_{\mu}\psi={\partial\psi\over\partial x^{\mu}}$ with respect to the coordinates
$x^{\mu}$ of the space-time point. According to the gauge principle,
this equation has to have the property
that if $\psi$ is a solution, then so is $e^{{iq\over\hbar}\chi}\,\psi$, 
whatever the function $\chi(x)$. This requires $\partial_{\mu}\psi$ to transform in the same way that $\psi$ does, as given by (2). This is impossible unless the notion of derivative is modified and replaced by what
 is called a \lq\lq covariant derivative'':
$$D_{\mu}\,\psi=\partial_{\mu}\psi-{iq\over\hbar}A_{\mu}\,\psi$$
with the understanding that in the evolution equation 
$\partial_{\mu}\psi$ always appears accompanied by
 $A_{\mu}$ in this combination. To understand the meaning
 of the newly introduced quantity $A_{\mu}$ it
 is necessary to see how it \lq\lq gauge transforms''.
Since we have found that the substitution (2) has to be  accompanied by
$$D_{\mu}\,\psi\rightarrow e^{{iq\over\hbar}\chi(x)}\,D_{\mu}\psi(x)$$
 a small calculation reveals that
$$A_{\mu}\rightarrow A_{\mu}+\partial_{\mu}\chi,\eqno(3)$$
the \lq\lq gauge transformation"
 of $A_{\mu}$.
Notice that this equation, (3), has the property of being universal 
in the sense that it is independent of $q$, the electric
 charge of the particle
initially considered. In other words, the argument works
for an array of particles with different electric charges,
even if they do not satisfy the quantisation condition (1).

Notice also that the \lq\lq curl" of $A_{\mu}$ is 
unaltered by the gauge transformation (3):
$$ F_{\mu\nu}\equiv
\partial_{\mu}\,A_{\nu}-\partial_{\nu}\,A_{\mu}\rightarrow F_{\mu\nu},\eqno(4)$$
as $\chi$ cancels out, by virtue of the irrelevance of the order of
differentiation with respect to independent space-time  variables.
 These final equations are very reminiscent
of Einstein's relativistic expression for the electromagnetic field tensor
incorporating both the electric and magnetic fields. In fact it can
be identified as precisely this. Then it follows that $A_{\mu}$ is simply
what is known as the gauge four-potential (composed of scalar and vector potentials).
In fact the above gauge ambiguity (3) specified by $\chi$ was well recognised
in the context of Maxwell's equations before the time of Planck's
proposal of quanta in 1900.

The identification of $F_{\mu\nu}$ defined this way as the
 Einstein-Maxwell electromagnetic field strength tensor 
implies that the physical  existence of this field is an inevitable consequence
of the gauge principle, the idea that physics is unaffected by the
local change of phase of wave functions given by (2). Furthermore
the structure of the covariant derivative above determines how
this field couples to any wave function and yields consequences
perfectly consistent with all that was known previously.

All of this argument works equally well in curved space-time.
 The result endows the electromagnetic field with a new, 
geometrical significance that elevates it to become on a par
with the gravitational field whose geometric meaning
was found so famously by Einstein. The definitive
version of this argument was found in 1929 by Hermann Weyl [1929],
better known as a pure mathematician. More information about the genesis of
the gauge principle, including the non-abelian version can be found in
the recent book by O'Raifeartaigh [1997]. This also includes a translation into English of the paper by Hermann Weyl just mentioned. Dirac was very quick
to exploit the ideas in a way that was competely novel, 
bringing into play ideas of  topology for the first time in physics.

But first I want to explain how an extension of the gauge principle 
yields one of the basic ingredients
of the standard model. Two successive gauge transformations (2)
can be combined to yield a third at each point $x$ of space-time
$$e^{{iq\over\hbar}\chi_1(x)}e^{{iq\over\hbar}\chi_2(x)}=
e^{{iq\over\hbar}(\chi_1(x)+\chi_2(x))}.$$
This means that the transformations at any point form
what is called a group, in this case the group of phase factors,
usually denoted $U(1)$ (unitary matrices with one row and column).
This particular group is abelian, that is the order of multiplication above
does not matter, as is evident from the formula.

If one considers an array of particles, for example quarks and leptons,
their wave functions can be arranged as a column
 vector with $N$ complex entries, say.  A natural way to consider of extending (2) to this situation would be the following:
$$\pmatrix{\psi_1(x)\cr\psi_2(x)\cr\psi_3(x)\cr.\cr.\cr\psi_N(x)\cr}\qquad
\Rightarrow\qquad
\pmatrix{D_{11}&D_{12}&D_{13}&.&.&D_{1N}\cr D_{21}&D_{22}&D_{23}&.&.&D_{2N}\cr
D_{31}&D_{32}&D_{33}&.&.&D_{3N}\cr
.&.&.&.&.&.\cr
.&.&.&.&.&.\cr D_{N1}&D_{N2}&D_{N3}&.&.&D_{NN}\cr}
\pmatrix{\psi_1(x)\cr\psi_2(x)\cr\psi_3(x)\cr.\cr.\cr\psi_N(x)\cr}$$
that is, to  premultiply  the column vector by a square matrix with $N$
 rows and columns and complex entries each dependent on the
 space-time point $x$.
 To preserve the total probability density this matrix should be unitary.
 Such substitutions again combine
 successively
 to form similar substitutions and so form a group at each point of space-time. As the order of multiplication now matters  very much, the group is said to be non-abelian.
The particular group formed in this way out of unitary matrices is called $U(N)$. If the determinants are constrained to equal unity a subgroup called  $SU(N)$ results.
These are examples of compact Lie groups and all other possibilities can be listed. But the examples just mentioned with $N$ taking the values $2$ and $3$ respectively suffice for the standard model.

 One of the \lq\lq devils in the detail'' that I mentioned earlier is the non-abelian nature of these groups.  The argument above  leading to the gauge theory can be repeated and the field strength is found to be no longer linear in the gauge potential, as in (4), since additional quadratic terms occur.
These mean that the equations of motion  are now  non-linear
and  so when, considered classically,  possess
 unexpected solutions of a soliton character.
 The process of quantisation is made more difficult
 and this is one of the problems with which Dirac was preoccupied later in his career. The theory of constraints covered in his advanced lecture course
was designed to aid the quantisation of theories with gauge symmetry and
was taken up successfully by later workers.

The purpose of my digression into the standard model,
 which was probably the main achievement
of the second half of the last century,
 was to make  clear the importance within it 
of the gauge principle. In fact, this is pervasive in all 
 modern thinking. Dirac used only the abelian, $U(1)$,
version, first enunciated in 1929, by Hermann Weyl. Within two years Dirac
 had succeeded in  exploiting it in the totally unexpected way that I shall now explain.

When the motion of an electrically charged particle is treated,
 ignoring quantum mechanics, it obeys Newton's equation of motion in a form including the Lorenz  force describing
 the effect of the electric and magnetic fields. This means that this equation, as well as the Maxwell equations,
involves the field strength $F_{\mu\nu}$ directly without any independent 
appearance of the gauge potentials $A_{\mu}$. This situation does not change if a natural modification is made to include particles carrying a
 magnetic charge.
 Indeed there appears to be a symmetry with respect to the  exchange of the roles of the electric and magnetic charges which is respected by the equations.

But the situation changes when quantum mechanics is introduced because of the way the
gauge potentials explicitly enter the covariant derivative of the wave
 function of the electrically charged particle, as explained above. Thus gauge potentials are needed to determine the evolution of the wave function and this makes it look as if
quantum mechanics is inconsistent with the presence of particles with non-zero magnetic charge, $g$. The reason is that the flux of magnetic field $\underline B$ through any surface, say a two-dimensional sphere, $S_2$, surrounding a magnetic particle is
$$\int_{S_2}\underline B.\underline{dS}=g.$$
On the sphere, $S_2$, $\nabla.\underline B$ vanishes as the magnetic charge is concentrated at one point, the position of the magnetically charged particle at its centre. This vanishing of $\nabla.\underline B$ is necessary for the
integrability of the equation 
 $\underline B=\nabla\wedge\underline A$, a special case of (4) with $\underline A$ denoting the vector potential, part of $A_{\mu}$.
But if this vector potential can be found all over the sphere then
Stokes' theorem implies that $g$ vanishes.  Dirac realised that this argument was wrong in a very subtle way. Instead of predicting $g=0$, he found that
$$qg=2\pi n\hbar, \qquad n=0,\pm1,\pm2\,\pm3\dots,
\eqno(5)$$
for $q$ being any electric charge and $g$ any magnetic charge.
 So long as a particle with non-zero magnetic charge $g$
 exists in nature, this value can be divided out of equation (5) 
to yield equation (1), the quantisation condition for electric charge which very much agrees with observation. The attractiveness of this result
evidently excited Dirac, but as he was aware, the price
 is the required existence of a magnetically charged particle. 
The repeated failure to observe isolated magnetic charge is what still 
 haunts the study of this subject.

In the seventy years since Dirac derived his quantisation condition (5) there has been a minor industry in refining his argument to take account
 of notions developed later in pure mathematics. 
I shall present the treatment of Wu and Yang [1975].

It is the first part of the argument that brings  topological ideas into play.
 Divide the sphere  surrounding the magnetic charge into two hemispheres,
north and south, which join on the equator. Unlike the sphere itself,
 each hemisphere
has the property that it can be distorted continuously 
to its appropriate pole, north or south, without tearing.
 This means that it is valid to integrate the equation 
$\underline B=\nabla\wedge\underline A$ to obtain  the
 vector potential $\underline A$ separately in each hemisphere,
 $\underline A^{\hbox{ \fiverm NORTH}}$
and $\underline A^{\hbox{ \fiverm SOUTH}}$, respectively, 
even though this cannot be done for the sphere as a whole
unless $g$ vanishes, as we saw. It follows that the two vector
 potentials $\underline A^{\hbox{ \fiverm NORTH}}$ and 
 $\underline A^{\hbox{ \fiverm SOUTH}}$
 must differ on the equator despite the fact
 $\nabla\wedge\underline A^{\hbox{ \fiverm NORTH}}$ and
 $\nabla\wedge \underline A^{\hbox{ \fiverm SOUTH}}$ coincide there,
 both equalling the magnetic field which is well defined all
 over the sphere, since it is
 a physically observable quantity. 
This means that $\underline A^{\hbox{ \fiverm NORTH}}$ and 
 $\underline A^{\hbox{ \fiverm SOUTH}}$ are related on the equator
 by a gauge transformation (3) which reads, in the present notation,
$$\underline A^{\hbox{ \fiverm NORTH}}= 
 \underline A^{\hbox{ \fiverm SOUTH}}+\nabla\chi.$$
The magnetic charge equals the sum of the magnetic fluxes
 out of the two hemispheres and each of these can be evaluated using Stokes'
theorem for each hemisphere. Each contribution is an integral of the 
appropriate gauge potential, northern or southern, over the boundary of the hemisphere which is, in both cases, the equator. In these integrals
the equator has to have a definite sense and these
must be relatively opposite since the sphere has no boundary,
being a closed surface. Accordingly
$$g=\int_{{\hbox{\fiverm NORTHERN}\atop \hbox{\fiverm HEMISPHERE}}}\underline B.\underline{dS}+
\int_{{\hbox{\fiverm SOUTHERN}\atop \hbox{\fiverm HEMISPHERE}}}\underline B.\underline{dS}$$
$$=\int_{\hbox{\fiverm EQUATOR}}(\underline  A^{\hbox{ \fiverm NORTH}}-  \underline A^{\hbox{ \fiverm SOUTH}}).\underline{dx}=\int_{\hbox{\fiverm EQUATOR}}\nabla\chi.\underline{dx}=\Delta\chi$$
using the gauge transformation and an integration by parts. 
Thus the magnetic charge, $g$, equals the jump, $\Delta\chi$, 
 in the value of the gauge function, $\chi$, as the equator 
is encircled once. 

The usefulness of this  relation is enhanced
when quantum mechanics is brought into play. Like the gauge potentials,
 the wave functions can be defined on each hemisphere separately 
 but not all over the sphere at once. On the equator they must differ by 
a gauge transformation (2), and,
 in order to tally with the connection between the two gauge
 potentials above,
$$\psi^{\hbox{ \fiverm NORTH}}(x)=e^{{iq\over\hbar}\chi(x)}\,
\psi^{\hbox{ \fiverm SOUTH}}(x)$$
As each wave function is single valued in its own hemisphere, both
are single-valued on the equator
 and it follows that the same also
 applies to the phase factor $e^{{iq\over\hbar}\chi(x)}$. Thus  
$e^{{iq\over\hbar}\Delta\chi}$ equals unity and this reduces to
Dirac's quantisation condition (5), when combined with the result above that $\Delta\chi=g$.

The nature of the subtle extra ingredient supplied by Dirac, 
beyond quantum mechanics and the gauge principle involves ideas
 of what is now known as topology. In particular Dirac's ideas
bring into play global topology which is the antithesis of locality,
involving the structure of spaces in the large rather than in the small.
The mathematical subject of topology 
 was in a rather primitive state in 1931
but has blossomed since then. With hindsight it can now be seen that
in 1931
Dirac had anticipated ideas of de Rham cohomology, Hodge theory,
fibre bundle theory and characteristic classes. I believe that Dirac was on friendly terms
 with his near contemporary, William Hodge, in the Cambridge of the 1930's.
Much later Hodge had a student, 
 Atiyah, who was to discover remarkable connections
between the Dirac equation and topology that have become an important part
of modern quantum field theory. Indeed topological ideas now permeate
that subject and even more modern string theory. This episode illustrates
the relationship of
Dirac's thinking to pure mathematics. Although his mathematical thought
was abstract and extremely effective, he seemed to dislike formalism
 {\it per se}, and it seems to me that the best description of
his  approach to pure mathematics was that it was  very DIY,
 that is \lq\lq do-it-yourself''.

I mentioned that the classical theory of particles carrying electric
 or magnetic charge possessed a formal symmetry with respect 
to the interchange of the electric and magnetic fields accompanied by a similar interchange of the two charges. The foregoing quantum argument seems to break the symmetry because of the preferred status of the electrically charged particles which are assigned wave functions while the magnetically charged particles are not, being treated as point objects. Despite this, the final result, Dirac's quantisation condition, (5), does respect the symmetry with respect
to interchange.

This prompts an  interesting question as to whether a more complete theory,
 containing
information concerning the structure and mass of these charged particles
 could also respect the symmetry. The advent of the standard model,
 or rather an extension known as a grand unified model,
 provides an arena in which an affirmative answer is possible.

 Just as quantum electrodynamics is embedded as a subtheory in
 the standard model, so is the latter embedded in a grand unified model,
 namely a gauge theory
with a simple Lie group as gauge group, $G_{GUT}$, and hence fewer
 free parameters. Corresponding to this the relevant gauge groups are embedded in each other
$$U(1)_Q\subset \hbox{\lq\lq$ U(2)\times SU(3)$''}\subset G_{GUT}.$$
$Q$ is the generator of the electromagnetic gauge group, $U(1)$, and hence
automatically a generator of the Lie algebra of the grand unified Lie group.
This fact has an important consequence: it can explain the quantisation 
of electric charge (1), apparently avoiding any need for magnetic charge.
 Indeed this was one of the original motives for considering such theories.

But it is necessary to ask how the direction of the electric charge
 is chosen out of the vector space of all directions in the Lie
 algebra of the grand unified group. The answer is
 that it is chosen by a special sort of Lorentz scalar field
that fails to vanish in the vacuum, unlike the other fields. This is known as a
\lq\lq Higgs'' field, and examples are already needed
 for other side-effects
such as the provision of  mass for the three weak gauge particles and
 the quarks whilst preserving full gauge invariance of the equations of motion.

 When combined with the non-linearities of the generalised Maxwell equations,
another effect of the Higgs field is the production
 of unexpected classical solutions, called solitons, that behave
 like stable, extended particles carrying magnetic charge,
 as 't Hooft [1974] and Polyakov [1974] independently demonstrated.
 Thus a theory such as this,  with  electric  charge quantisation inbuilt, automatically
produces magnetically charged particle states. Furthermore,
 Dirac's quantisation condition (5) is seen to hold, arising in a way that
has to to with the topology of the map provided by the Higgs field. 
Thus the survival of Dirac's argument  in a modern context
demonstrates its pervasiveness.

In special cases at least, such as when $G_{GUT}\equiv SO(3)$,
 the forementioned interchange symmetry between electric
 and magnetic charge does hold good, even
when account is taken of expressions for the particle masses yielded by the Higgs effect [Montonen and Olive 1978]. As the single particle states satisfy the Dirac quantisation condition (5) with $n=2$, this interchange reduces to
$$q\rightarrow g={4\pi\hbar\over q}, \eqno(6)$$
and is now manifestly quantum in nature. More remarkably there exist an infinite number of distinct, stable, single particle states carrying both electric and magnetic charge, and called \lq\lq dyons'' as a consequence. The theory 
exhibits a symmetry when these are  permuted suitably. At the same time a new dimensionless
 parameter $\theta$ enters in addition to the fine structure constant $\alpha={q^2\over4\pi\hbar}$ and it turns out to be  useful to unify
the two by  defining the following complex parameter:
$$\tau={\theta\over2\pi}+{i\over\alpha}.\eqno(7)$$
It can be shown [Sen 1994] that the allowed permutations of dyonic particle states have
 the effect
$$\tau\rightarrow {a\tau+b\over c\tau+d},\qquad a,b,c,d\hbox{ integers, }\qquad ad-bc=1.\eqno(8)$$
These fractional linear transformations form an infinite discrete group called the modular group,
 first studied by pure mathematicians in the 19th century. They realised that it had the important property of maintaining the positivity of the imaginary part of $\tau$, physically the inverse of the fine structure constant 
according to expression (7),
 and so intrinsically positive.

To understand the transformations (8), first note 
that when $\theta=0$ the transformation
 $\pmatrix{a&b\cr c&d\cr}=\pmatrix{0&-1\cr 1&\,\,\,0\cr}$ yields the original interchange (6), whilst $\pmatrix{a&b\cr c&d\cr}=\pmatrix{1&1\cr0&1\cr}$ simply increases $\theta$ by $2\pi$, signifying that it is an angular variable as the notation suggests. Repetition of these two transformations yields all of the modular group.

According to the ideas of perturbation theory, it is possible to make good approximate calculations when the fine structure constant $\alpha$ is small but not otherwise. So the behaviour of the theory when $\alpha$ is large is considered
inaccessible. But the transformation (6) interchanges large and small values of $\alpha$ and so relates the unknown to the calculable. At first sight this is difficult to believe but a fair number of checks have confirmed the plausibility of this idea and it has become a matter of faith in much current research.

Another advantageous special condition on the spontaneously broken gauge theory is that it should display supersymmetry and this is easy to arrange. One consequence is that many of the divergences that troubled Dirac automatically cancel. Furthermore superstring theory contains such theories as a limiting case and, as a result, the ideas extend rather naturally to that framework. Making all that
I have said more precise, watertight and wide-ranging is now a major priority in current research.

The influential, or even, prophetic nature of Dirac's thinking is apparent in the above story. There are also examples in his later work that turned out to be likewise prophetic and/or influential. He was the first to advocate
use of expressions in quantum electrodynamics that
 are akin to what are now called Wilson loops [Dirac 1955b].
In his bubble theory
of the electron [Dirac 1962] he anticipated the Nambu-Goto  action for the string and the later membrane action relevant to M-theory. His wave-front approach to quantum field theory [Dirac 1949] 
introduced the light-cone quantisation procedure applied so
 often subsequently. His analysis of the equal-time  commutation relations 
of the energy-momentum tensor [Dirac 1955a] blossomed into the conformal field theory
 treatment of the critical exponents in the theory of second order phase
 transitions.
During the 1950's and 1960's one major preoccupation of Dirac
 was the quantisation of constrained dynamical sytems as a step towards the
quantisation of systems with gauge symmetry such as Einstein's theory
 of general relativity. Although the latter problem  remains unsolved,
his methods have proved valuable in the context of non-abelian gauge theories
and hence in the standard model
 [Dirac 1950]. As I mentioned earlier he lectured on this subject
 in Cambridge and also wrote a book, newly republished [Dirac 1964].
  The study of the 
strange representation of the anti-de-Sitter group [Dirac 1963] has
burgeoned into the study of the metaplectic representation and something called Howe duality, quite possibly related to the electromagnetic duality explained above.

I have tried to explain briefly the pervasiveness of Dirac's thought in modern theoretical physics by a few examples and, in particular, by reference to the
development of the  standard model and the era that followed. The latter example also illustrates the impact of the role played by Hermann Weyl, famous as a 
pure mathematician, but less recognised as a theoretical physicist.

Perspectives on the status of past scientific contributions
are liable to change in the light of improvements in our
 own understanding following from
 general progress in the subject. I believe that the new perspective
 gained since the establishment of the standard model has enhanced Dirac's
 reputation.
Granted that he was one of the towering figures behind the development
of quantum theory and relativistic quantum field theory, we can 
 now see how his other ideas that were
 initially less appreciated 
 are finally achieving similar degrees of
appreciation and recognition for their
 ingenuity, correctness and relevance.

\medskip
Finally I would like to thank Bruce  Gordon for organising such a pleasant meeting.

\bigskip
\def\ni{\noindent}
\noindent {\bf References}
\bigskip

\ni P.~A.~M.~Dirac, \lq\lq The quantum theory of the emission and absorption of radiation'',  Proc.\ Roy.
\ Soc. \ London \ A {\bf 114} (1927) 243-265.

\ni P.~A.~M.~Dirac, \lq\lq The quantum theory of the electron'', Proc.\ Roy.
\ Soc. \ London \ A {\bf 117} (1928) 610-624.

\ni P.~A.~M.~Dirac, \lq\lq Quantised singularities in the
 electromagnetic field'', Proc.\ Roy.\ Soc.\ Lond.\ A{\bf 133} (1931) 60-72.

\ni P.~A.~M.~Dirac,
``Forms Of Relativistic Dynamics,''
Rev.\ Mod.\ Phys.\  {\bf 21} (1949) 392-399.

\ni P.~A.~M.~Dirac,
``Generalized Hamiltonian Dynamics,'' 
Can.\ Journ.\ Math.\ {\bf 2} (1950) 129-148 and 
Proc.\ Roy.\ Soc.\ Lond.\ A {\bf 246} (1958) 326.

\ni P.~A.~M.~Dirac, \lq\lq The stress tensor in field dynamics'',
 Nuovo Cim. (10) {\bf 1} (1955) 16-36.

\ni P.~A.~M.~Dirac,
``Gauge Invariant Formulation Of Quantum Electrodynamics,''
Can.\ J.\ Phys.\  {\bf 33} (1955) 650-660.

\ni P.~A.~M.~Dirac,
``An Extensible Model Of The Electron,''
Proc.\ Roy.\ Soc.\ Lond.\ A {\bf 268} (1962) 57-67.

\ni P.~A.~M.~Dirac,
``A Remarkable Representation Of The 3+2 De Sitter Group,''
J.\ Math.\ Phys.\  {\bf 4}, (1963)  901-909.

\ni P.~A.~M.~Dirac, \lq\lq Lectures on Quantum mechanics'', Belfer
Graduate School of Science, Yeshiva Univ, 1964 and Dover 2001.

\ni  G.~'t Hooft,
``Magnetic Monopoles In Unified Gauge Theories,''
Nucl.\ Phys.\ B {\bf 79} (1974) 276-284.

\ni C.~Montonen and D.~Olive, ``Magnetic Monopoles As Gauge Particles?,''
Phys.\ Lett.\ B {\bf 72} (1977) 117-120.

\ni D.~I.~Olive,\lq\lq The relativistic electron'' in \lq\lq Electron,
a centenary volume'', edited by M.~Springford, CUP 1997.

\ni L.~O'Raifeartaigh, \lq\lq The dawning of the gauge principle'',
Princeton 1997.

\ni A.~M.~Polyakov,
``Particle Spectrum In Quantum Field Theory,''
JETP Lett.\  {\bf 20} (1974) 194-195
[Pisma Zh.\ Eksp.\ Teor.\ Fiz.\  {\bf 20} (1974) 430].

\ni A.~Sen, \lq\lq Dyon-monopole bound states,
self-dual harmonic forms on the multi-monopole moduli space, and
$SL(2,Z)$ invariance in string theory", Phys. Lett.\ {\bf B329} (1994), 217-221.

\ni H.~Weyl, Z. Physik {\bf 56} (1929) 56.

\ni T.~T.~Wu and C.~N.~Yang, \lq\lq Concept of non-integrable phase factors
and global formulation of gauge fields'',
Phys. Rev. {\bf D12} (1975) 3845-3857.

\bye